\documentclass[twocolumn]{jpsj3}

\usepackage{color}
\usepackage{bm}

\title{%
Unified Description of Dirac Electrons on a Curved Surface
of Topological Insulators
}

\author{%
Yositake {\sc Takane} and Ken-Ichiro {\sc Imura}
}

\inst{%
Department of Quantum Matter, Graduate School of Advanced Sciences of Matter,\\
Hiroshima University, Higashihiroshima, Hiroshima 739-8530, Japan
}

\recdate{ \hspace{50mm} }

\abst{%
Existence of a protected surface state described by a massless Dirac equation
is a defining property of the topological insulator.
Though this statement can be explicitly verified on an idealized flat surface,
it remains to be addressed to what extent it could be general.
On a curved surface, the surface Dirac equation is modified by 
the spin connection terms.
Here, in the light of the differential geometry,
we give a general framework for constructing the surface Dirac equation
starting from the Hamiltonian for bulk topological insulators.
The obtained unified description clarifies the physical meaning of
the spin connection.
}

\kword{%
strong topological insulator, Dirac equation, surface state, spin connection
}

\begin{document}
\sloppy
\maketitle

\section{Introduction}

Low-energy electron states on the surface of a topological insulator
is protected by the time reversal symmetry,
which is encoded as (a nontrivial value of)
the $\mathbb{Z}_2$ topological index.~\cite{fu,moore,roy}
Electrons in these surface states obey a massless Dirac equation,
and possess a linear energy dispersion forming a gapless conic structure,
i.e., a Dirac cone in the reciprocal space.
Such a conic dispersion has been observed experimentally
by angle-resolved photoemission spectroscopy
measurements.~\cite{ARPES0,ARPES1,ARPES2,ARPES3,ARPES4}

The existence of a single Dirac cone
on an ideally flat large cleaved surface of
a three-dimensional (3D), strong topological insulator
is thus incontrovertible,~\cite{hasan-kane}
while it is less clear how the low-energy electrons behave
on {\it curved} surfaces.~\cite{lee,zhang1,zhang2}
Realistic surfaces of the topological insulator have
imperfections, such as terraces and islands,~\cite{takane,yy}
that have been also experimentally observed.~\cite{STM}
Samples of a finite volume have a closed surface as a whole,
or a set of facets.
The corner of two facets are sometimes better described as a
curved single surface.~\cite{takane}
In addition to such natural occurrence, the role of the curvature
becomes particularly important
in artificially fabricated topological insulator nanostructures
such as nanowires~\cite{nanowire}
and nanoparticles.~\cite{spherical}

The behavior of low-energy electrons 
has been studied theoretically, on a curved surface of
e.g., cylindrical~\cite{zhang1,zhang2,egger,bardarson,imura1,imura3}
or spherical~\cite{lee,parente,spherical} form.
In case of the cylindrical geometry,
a remarkable feature is the appearance of a finite-size energy gap
separating the upper and lower Dirac cones of surface electron states.
In refs.~\citen{zhang1,zhang2,egger,bardarson,imura1,imura3},
the origin of this gap has been much discussed;
the opening of the gap was attributed to
half-odd integral quantization of the orbital angular momentum
($L_z$ in a cylinder pointed to the $z$-axis).
The quantization is half-shifted by what is often called
the ``Berry's phase $\pi$''.
Most schematically, this interpretation may be sketched as follows:
first, the real spin of a surface electron is constrained
onto the tangential plane at any point on a curved surface;
this feature is often referred to as ``spin-to-surface locking''.
As a consequence of this locking,
when the electron goes around the cylinder,
the surface curvature induces ``spin connection''~\cite{spinc}
that sums up to the Berry's phase $\pi$.

In spite of the accumulation of such studies,
we have so far not reached the unified description
of the low-energy electrons on curved surfaces.
When existence of a protected surface state is suggested
by the bulk-boundary correspondence,~\cite{bbc1,bbc2}
it is natural to expect that low-energy electron states are described
by a Dirac-type equation even on curved surfaces,
while an explicit demonstration of this has been lacking.
Note that the bulk-boundary correspondence ensures
on a flat surface
the existence of a protected {\it gapless} surface state described
by a massless Dirac Hamiltonian,
while on curved surfaces one generally expects that
such an effective surface Dirac Hamiltonian
will be modified by spin connection terms.

Let us try to give a brief overview of the present status of
theoretical studies dealing with a curved surface of topological insulators.
This issue has been addressed
mainly in cylindrical~\cite{zhang1,zhang2,egger,bardarson,imura1,imura3}
and spherical~\cite{lee,parente,spherical} systems
by employing two different types of approaches,
with the exception of refs.~\citen{takane} and \citen{imura4}
that have treated the case of hyperbolic surfaces.
The first group of studies is based on the two-dimensional (2D)
Dirac equation for a flat surface,
and takes account of the curved nature of the surface by a coordinate
transformation.~\cite{lee,zhang1,zhang2,egger,bardarson,parente}
Certainly, the resulting curved surface Dirac theory can be applicable 
to an arbitrary curved surface in this approach,
but for a clear reason that it assumes a 2D Dirac theory from the outset,
it fails to answer the question whether
the low-energy electrons obey indeed a Dirac-type equation.
The drawback of this approach is that it ignores
the 3D nature of the original problem.
If one recalls the importance of bulk-boundary correspondence
in the conceptual foundation of the topological insulator,
it may not be surprising that such an approach turns out to keep
only ``half of the information''.

In the second group of approaches, on contrary,
such a difficulty is well overcome.
There, one starts from a 3D bulk Hamiltonian
and derives an effective 2D theory for a given curved surface
by the use of the $\bm k \cdot \bm p$
approximation.~\cite{imura1,takane,spherical,imura3,imura4}
By its construction such an approach
takes well account of the 3D nature of the problem.
In the remainder of this paper, we extend this type of analyses
so as to describe the surface state of a topological insulator
of an arbitrary shape.
As a result, we are led to the unified description of
such a surface state on arbitrary curved surfaces.

Construction of the general framework may also help to clarify
the following issues related to the curved surface Dirac theory.
One is on the origin of the Berry's phase $\pi$
in the context of spin-to-surface locking.
As already mentioned, the Berry's phase $\pi$ is recognized
as a consequence of $\pm 2\pi$ rotation of the real spin caused by
the spin-to-surface locking.~\cite{zhang1,zhang2,egger,bardarson,imura1}
This locking, however, breaks down, for example,
in spherical topological insulators,~\cite{spherical}
since any spin configurations under the complete spin-to-surface locking
cannot satisfy the periodic boundary condition for the sphere
without singularities.
This fact raises a natural question: is the spin-to-surface locking
essential in the appearance of the Berry's phase $\pi$?
Note here that a finite-size energy gap associated with the spin connection
also appears in the spherical geometry.~\cite{lee,parente,spherical}

Another issue that will be addressed in the paper
concerns the ``physical'' interpretation of the spin connection.
The spin connection appears in the curved surface Dirac theory
as a fictitious vector potential in the effective Hamiltonian.
Though mathematically well-defined,
encoding the information on the curved nature of the surface,
its physical origin is somewhat mysterious,
especially for those researchers in the condensed matter community.
Here, in this paper we attempt an accessible presentation of this issue.

In the next section we introduce a set of curvilinear coordinates
adapted for describing an arbitrary curved surface.
In \S3, we present a general framework to derive
the effective surface Hamiltonian for the low-energy electrons
within the $\bm k \cdot \bm p$ approximation.
We show that the effective Hamiltonian is indeed expressed
in a generalized Dirac form with the spin connection.
The role of the spin connection is
clarified and given an intuitive interpretation.
In \S4, we discuss the boundary condition for a spinor wave function
in spatially periodic systems, and identify
the precise origin of the Berry's phase $\pi$.
In \S5, our framework is applied to two known examples, 
namely to the case of cylindrical and spherical geometries,
to demonstrate how it works.
The last section is devoted to summary and discussion.
We set $\hbar = 1$ throughout this paper.

\section{Basic formulations}

Before introducing a curved surface and a set of proper (curvilinear)
coordinates for describing it,
let us first specify what is in the bulk.
We start with the following bulk effective Hamiltonian
for a (3D, isotropic, strong) topological insulator
in the continuum limit:~\cite{liu,shan}
$H_{\rm bulk} = m(\mib{p})\tau_{z} + A\tau_{x}(\mib{\sigma}\cdot\mib{p})$,
where $\mib{p}=-{\rm i}\nabla$ and
$m({\mib p}) = m_0 + m_2\mib{p}^{2}$ is the mass term.
The two types of Pauli matrices $\mib{\sigma} =(\sigma_x, \sigma_y, \sigma_z)$
and $\mib{\tau} =(\tau_x, \tau_y, \tau_z)$ represent, respectively,
the real and the orbital spin degrees of freedom.
If the ordinary matrix representation of $\mib{\tau}$ is used,
$H_{\rm bulk}$ is expressed as
\begin{align}
       \label{matH}
   H_{\rm bulk}
   = \left[
   \begin{array}{cc}
             m({\mib p}) & A(\mib{\sigma}\cdot\mib{p}) \\
             A(\mib{\sigma}\cdot\mib{p}) & -m({\mib p})
           \end{array}
     \right] .
\end{align}
Throughout the paper
we assume that mass parameters are chosen such that
$m_{0}>0$ and $m_{2}<0$.
When this is the case, 
the existence of a protected surface state is ensured
by the bulk-boundary correspondence.
On a flat surface, 
the surface state exhibits a gapless spectrum,
described by a massless Dirac Hamiltonian.
On a curved surface, the Dirac equation is modified by the
spin connection terms, as we demonstrate below.

\begin{figure}[tbp]
\begin{center}
\includegraphics[height=4.5cm]{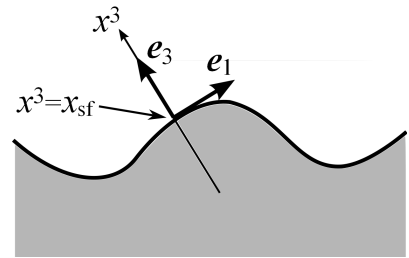}
\end{center}
\caption{Curved surface of a 3D topological insulator: 
cross section in the plane
spanned by ${\mib e}_{1}$ and ${\mib e}_{3}$.
}
\end{figure}
Let us consider a curved surface of the sample described by
two coordinates $(x^{1},x^{2})$ as $X^{\alpha} = f^{\alpha}(x^{1},x^{2})$,
where $\mib{X}=(X^{1},X^{2},X^{3})$ represents the position of
an arbitrary point on the surface
in the 3D Cartesian coordinates (see Fig.~1).
Depending on the geometry (e.g., closed vs. open surfaces, etc.),
the coordinate $x_i$ represents either a linear (non-cyclic) or
a cyclic coordinate.
For simplicity we focus on samples whose entire surface is described by
a single set of functions $\{f^{\alpha}\}$,
although with little modification our argument can be extended to cases
where the entire surface is wrapped by several patches
and a separate set of functions is needed to describe each patch.
We introduce a set of curvilinear coordinates 
suitable for analyzing 
low-energy electron states of the topological insulator
localized in the vicinity of its surface.
As surface electron states have a finite penetration depth $\lambda$,
we require that the set of curvilinear coordinates are well-defined
only in the surface region of width on the order of $\lambda$.

Let ${\mib e}_{1}$ and ${\mib e}_{2}$ be the two tangent vectors defined by
\begin{align}
     \label{eq:mib-e-def}
  {\mib e}_{i} = \frac{\partial \mib{X}}{\partial x^{i}} .
\end{align}
Note that ${\mib e}_{1}$ and ${\mib e}_{2}$ are not necessarily
orthogonal with each other nor normalized to be unity.
Let ${\mib e}_{3}$ be the unit normal vector defined by
\begin{align}
  {\mib e}_{3} = \frac{{\mib e}_{1}\times{\mib e}_{2}}
                      {|{\mib e}_{1}\times{\mib e}_{2}|} ,
\end{align}
which for simplicity is assumed to be outward normal to the surface.
We introduce the third (perpendicular) coordinate $x^{3}$
along the straight line designated by ${\mib e}_{3}$,
and set $x^{3} = x_{\rm sf}$ just on the surface.
Let us consider fictitious internal surfaces obtained
by varying $(x^{1},x^{2})$ at fixed values of $x^{3}$
satisfying $x_{\rm sf} \ge x^{3} \gtrsim x_{\rm sf}-\lambda$.
These surfaces are well-defined over the entire sample as long as
the smallest value of local radius of curvature at $x^{3} = x_{\rm sf}$
is much longer than $\lambda$.
Let $f^{\alpha}(x^{1},x^{2},x^{3})$ be the function
that describes the internal surface at $x^{3}$.
In terms of this we define ${\mib e}_{1}$ and ${\mib e}_{2}$
similar to eq.~(\ref{eq:mib-e-def}).
Now ${\mib e}_{1}$ and ${\mib e}_{2}$ become functions of
$(x^{1},x^{2},x^{3})$, while ${\mib e}_{3}$ is independent of $x^{3}$.
We employ $(x^{1},x^{2},x^{3})$ as the curvilinear coordinates with
the basis vectors
\begin{align}
 {\mib e}_{1} & = {\mib e}_{1}(x^{1},x^{2},x^{3}) ,
      \\
 {\mib e}_{2} & = {\mib e}_{2}(x^{1},x^{2},x^{3}) ,
      \\
 {\mib e}_{3} & = {\mib e}_{3}(x^{1},x^{2}) .
\end{align}

Let us introduce ${\mib e}^{i}$
that satisfies ${\mib e}_{i}\cdot{\mib e}^{j}=\delta_{ij}$.
We also introduce the metric tensors in a symmetric bilinear form
defined by $g_{ij}\equiv{\mib e}_{i}\cdot{\mib e}_{j}$
and $g^{ij}\equiv{\mib e}^{i}\cdot{\mib e}^{j}$,
which satisfy $g_{ik}g^{kj}=\delta_{ij}$.
Here and hereafter we use the convention that
a repeated index, such as $i$, $j$, and $k$, should be summed over $1, 2, 3$.
Obviously, $g_{33}=g^{33}=1$ as $|\mib{e}_{3}|=1$,
and $g_{13}=g_{23}=g^{13}=g^{23} = 0$
as ${\mib e}_{1} \bot {\mib e}_{3}$ and ${\mib e}_{2} \bot {\mib e}_{3}$.
In this coordinate system, an infinitesimal volume element is given by
${\rm d}V ={\rm d}S{\rm d}x^{3}$ with 
\begin{align}
  {\rm d}S = \sqrt{{\mathcal G}}{\rm d}x^{1}{\rm d}x^{2}
\end{align}
being an infinitesimal area element, where
\begin{align}
   \mathcal{G}(x^{1},x^{2},x^{3})
   \equiv {\rm det}\{g_{ij}\}
   = g_{11}g_{22}-g_{12}^{2} .
\end{align}
Within the curvilinear coordinates presented above,
the 3D Cartesian coordinates $(X^{1},X^{2},X^{3})=(x,y,z)$
and the differential operators with respect to them
are represented as $X^{\alpha} = e_{i\alpha}x^{i}$
and $\partial/\partial{X^{\alpha}} = {e^{i}}_{\alpha}\partial_{i}$,
where $\partial_{i} \equiv \partial/\partial{x^{i}}$.
For later convenience we decompose the Laplacian into two parts as
$\nabla^{2} = \Lambda_{\perp}+\Lambda_{\parallel}$, where
\begin{align}
    \label{eq:Lap-perp}
  \Lambda_{\perp}
  & = \frac{1}{\sqrt{\mathcal G}}\partial_{3}
      \left(\sqrt{\mathcal G}\partial_{3} \right) ,
      \\
    \label{eq:Lap-parall}
  \Lambda_{\parallel}
  & = \sum_{i,j=1}^{2}
      \frac{1}{\sqrt{\mathcal G}}\partial_{i}
      \left(\sqrt{\mathcal G}g^{ij}\partial_{j} \right) .
\end{align}

Finally we present a set of equations which describe
the spatial variation of ${\mib e}_{i}$ against $x^1$ and $x^2$.
The tangent vectors obey the Gauss equation,
\begin{align}
     \label{eq:Gauss}
  \partial_{j}{\mib e}_{i}
  = \sum_{k=1}^{2}\Gamma_{ij}^{k}{\mib e}_{k} + b_{ij}{\mib e}_{3} ,
\end{align}
where
\begin{align}
  b_{ij}
  & = - (\partial_{j}{\mib e}_{3})\cdot{\mib e}_{i} ,
       \\
  \Gamma_{ij}^{k}
  & = \frac{1}{2}g^{kl}
      \left( \partial_{j}g_{li}+\partial_{i}g_{lj}-\partial_{l}g_{ij}
      \right) .
\end{align}
The unit normal vector obeys the Weingarten equation,
\begin{align}
    \label{eq:Weingarten}
  \partial_{j}{\mib e}_{3}
  = - \sum_{i=1}^{2}{b_{j}}^{i}{\mib e}_{i} ,
\end{align}
where
\begin{align}
  {b_{j}}^{i}
  = \sum_{k=1}^{2}g^{ik}b_{kj}
  = - (\partial_{j}{\mib e}_{3})\cdot{\mib e}^{i} .
\end{align}
These equations are used in Appendices B and C in deriving,
or simplifying, matrix elements of the effective Hamiltonian.

\section{The effective Hamiltonian in a Dirac form}

Let us now derive the effective Hamiltonian
on the curved surface
specified by the curvilinear coordinates $(x^{1},x^{2},x^{3})$.
The derivation consists of four steps.

In the first step, we rewrite eq.~(\ref{matH}) in terms of
the curvilinear coordinates, and then divide it into 
components, either perpendicular or parallel to
the local tangent of curved surface spanned by
${\mib e}_{1}$ and ${\mib e}_{2}$.
The former describes penetration of the surface wave functions into the bulk, 
while the latter determines
low-energy properties of the surface states.
As a result, $H_{\rm bulk}$ is decomposed as
$H_{\rm bulk} = H_\perp + H_{\parallel}$ with
\begin{align}
       \label{H_perp}
   H_{\perp}
 & = \left[
       \begin{array}{cc}
         m_{0}-m_{2}\Lambda_{\perp} & -{\rm i}A\sigma^{3}\partial_{3} \\
         -{\rm i}A\sigma^{3}\partial_{3} & -m_{0}+m_{2}\Lambda_{\perp}
       \end{array}
     \right] ,
           \\
       \label{delta_H}
   H_{\parallel}
 & = \left[
       \begin{array}{cc}
          -m_{2}\Lambda_{\parallel}
           & -{\rm i}A\sum_{i=1}^{2}\sigma^{i}\partial_{i} \\
          -{\rm i}A\sum_{i=1}^{2}\sigma^{i}\partial_{i}
           & m_{2}\Lambda_{\parallel}
       \end{array}
     \right] ,
\end{align}
where we have used $\mib{\sigma}\cdot\nabla = \sigma^{i}\cdot\partial_{i}$
with
\begin{align}
  \sigma^{i} \equiv {\mib e}^{i}\cdot\mib{\sigma} .
\end{align}

In the second step, we solve the eigenvalue equation
$H_\perp |\psi \rangle
= E_\perp |\psi \rangle$,~\cite{liu,shan,imura1,spherical}
associated with the perpendicular part (\ref{H_perp}),
to find the two basis eigenstates $|\pm\rangle$
for constructing the surface effective Hamiltonian.
The appropriate boundary condition for $|\psi \rangle$
is $|\psi (x^{3}=x_{\rm sf})\rangle=\mib 0$.
That is, all four components of the wave function
$|\psi\rangle$ vanish on the surface at $x^{3} = x_{\rm sf}$.
As the simplest approximation,
we replace $\Lambda_{\perp}$ in $H_{\perp}$ with
\begin{align}
      \label{eq:Lam-perp-appr}
  \Lambda_{\perp}
  & = \partial_{3}^{2}
      + \frac{1}{2}\langle\partial_{3}{\rm ln}{\mathcal G}\rangle\partial_{3}
      \nonumber \\
  & \equiv \partial_{3}^{2}
           + \Delta(x^{1},x^{2}) \partial_{3} ,
\end{align}
where the definition of the average $\langle \cdots \rangle$ over $x^{3}$
is given below [see eq.~(\ref{eq:def-ave})].
Then, we can show that the eigenvalue equation
has surface solutions of the damped form,
$|\psi \rangle = {\rm e}^{\kappa (x^{3}-x_{\rm ef})} |u \rangle$.
Here $\kappa$ characterizing the penetration depth $\lambda$
is determined by ${\rm det}\{M_{\perp}\}=0$ with
\begin{align}
  M_{\perp}
  = \left[
      \begin{array}{cc}
         m_{0}-m_{2}\zeta-E_{\perp}
         & -{\rm i}A\kappa\sigma^{3} \\
         -{\rm i}A\kappa\sigma^{3}
         & -m_{0}+m_{2}\zeta-E_{\perp}
      \end{array}
    \right] ,
\end{align}
where $\zeta = \kappa^{2}+\Delta\kappa$.
Let us introduce the two eigenvectors $\mib{n}_{\pm}$ of $\sigma^{3}$.
They satisfy
\begin{align}
    \label{eq:eigen-sigma3}
 \sigma^{3}\mib{n}_{\pm} = \pm\mib{n}_{\pm} ,
\end{align}
and are regarded as local spin quantization axis.
$\mib{n}_{\pm}$ points
in the $\pm{\mib e}_{3}$ direction
if the spin axes $(s_{x},s_{y},s_{z})$ are identified with
the 3D Cartesian coordinates $(x,y,z)$.
That is, $\mib{n}_{+}$ ($\mib{n}_{-}$) is outward (inward)
normal to the tangential plane at $(x^{1},x^{2})$ on the surface.
In terms of the $4 \times 4$ unitary matrix $U$ defined by
$U={\rm diag}\{u,u\}$ with $u=(\mib{n}_{+},\mib{n}_{-})$,
we can partially diagonalize $M_{\perp}$ in the real spin space as
\begin{align}
  U^{\dagger}M_{\perp}U
  = \left[
      \begin{array}{cc}
         m_{0}-m_{2}\zeta-E_{\perp}
         & -{\rm i}A\kappa\sigma_{z} \\
         -{\rm i}A\kappa\sigma_{z}
         & -m_{0}+m_{2}\zeta-E_{\perp}
      \end{array}
    \right] .
\end{align}
This implies
\begin{align}
  {\rm det}\{M_{\perp}\}
  = [E_{\perp}^{2}+(A\kappa)^{2}-(m_{0}-m_{2}\zeta)^{2}]^{2}=0,
\end{align}
i.e., among the four solutions for $\kappa$
there are two pairs of solutions that are different only in their sign.
This means that among the four solutions of the eigenvalue problem
two are exponentially decreasing functions 
toward the interior (bulk) of the sample (i.e., $\kappa > 0$),
albeit the remaining two solutions being exponentially increasing.
By superposing two of such exponentially decreasing (i.e., normalizable)
solutions with $\kappa=\kappa_{\pm}$,
we construct a general solution localized near the surface as
\begin{align}
  |\psi \rangle
  = {\rm e}^{\kappa_{-}(x^{3}-x_{\rm sf})}|u_{-}\rangle
   -{\rm e}^{\kappa_{+}(x^{3}-x_{\rm sf})}|u_{+}\rangle .
\end{align}
The boundary condition $|\psi (x^{3}=x_{\rm sf})\rangle=\mib 0$
holds only when
$|u_{+}\rangle = |u_{-}\rangle$ for $\kappa_{+} \neq \kappa_{-}$. 
As shown in Appendix A, we see that this results in
the zero-energy condition $E_{\perp} = 0$ and
\begin{align}
     \label{eq:kappa-def}
  \kappa_{\pm}(x^{1},x^{2})
  = \frac{{\cal A} \pm \sqrt{{\cal A}^{2}+4m_{0}m_{2}}}
         {-2m_{2}}
\end{align}
with ${\cal A} \equiv A+m_{2}\Delta$.
We also find that two basis eigenstates, $|+ \rangle$
and $|- \rangle$, for $H_\perp$ with $E_{\perp} = 0$ are given by
\begin{align}
       \label{eq:pm-def}
   |\pm \rangle(x^{1},x^{2},x^{3})
   = \rho(x^{3};x^{1},x^{2}) |\pm \rangle\rangle(x^{1},x^{2})
\end{align}
with
\begin{align}
       \label{eq:pm_vec-def}
  |\pm \rangle\rangle(x^{1},x^{2})
   = \frac{1}{\sqrt{2}}
     \left[
       \begin{array}{c}
         \mib{n}_{+} \\
         \mp{\rm i}\mib{n}_{-}
       \end{array}
     \right]
\end{align}
and
\begin{align}
  \rho(x^{3};x^{1},x^{2})
    = \frac{1}{\sqrt{c}}
      \left( {\rm e}^{\kappa_{-}(x^{3}-x_{\rm sf})}
            -{\rm e}^{\kappa_{+}(x^{3}-x_{\rm sf})}
      \right) ,
\end{align}
where $c(x^{1},x^{2})$ is a normalization constant.
The $x^{1}$- and $x^{2}$-dependences of $\rho$ arise from
$c$ and $\kappa_{\pm}$.

In the third step we derive the effective surface Hamiltonian
within the $\bm k \cdot \bm p$ approximation.
The following derivation is based on the observation that
any surface state $|\Psi \rangle$ can be represented
as a linear combination of $|+ \rangle$ and $|- \rangle$
with the amplitude respectively specified by
$\tilde{\alpha}_+$ and $\tilde{\alpha}_-$, i.e.,
\begin{align}
     |\Psi\rangle=\tilde{\alpha}_+ |+\rangle + \tilde{\alpha}_-|- \rangle .
\end{align}
Within the $\bm k \cdot \bm p$ approximation,
the effective surface Hamiltonian $\tilde{\cal H}_{\rm eff}$
for the two-component spinor
$\tilde{\mib\alpha} =\, ^{t}(\tilde{\alpha}_{+}, \tilde{\alpha}_{-})$
is given by
\begin{align}
  \tilde{\cal H}_{\rm eff}
  = \left[
      \begin{array}{cc}
        \langle +| H_{\parallel} |+ \rangle &
        \langle +| H_{\parallel} |- \rangle \\
        \langle -| H_{\parallel} |+ \rangle &
        \langle -| H_{\parallel} |- \rangle
      \end{array}
    \right].
\end{align}
Here, each matrix element is expressed by
\begin{align}
      \label{eq:mat-H_eff}
  \langle \sigma| H_{\parallel} |\sigma' \rangle
  = \int_{x_{\rm sf}-l_{\rm c}}^{x_{\rm sf}}{\rm d}x^{3}
    \sqrt{{\cal G}}
    \rho\langle\langle \sigma | H_{\parallel} |\sigma' \rangle\rangle\rho ,
\end{align}
where $l_{\rm c}$ is the cutoff length being much longer than
the penetration depth $\lambda$.
Note that the factor $\sqrt{{\cal G}}$ reflects the fact that
${\rm d}V = \sqrt{{\cal G}}{\rm d}x^{1}{\rm d}x^{2}{\rm d}x^{3}$.
Accordingly, we set the normalization constant as
\begin{align}
   c(x^{1},x^{2})
 = \int_{x_{\rm sf}-l_{\rm c}}^{x_{\rm sf}}{\rm d}x^{3}
   \sqrt{{\cal G}}
   \left( {\rm e}^{\kappa_{-}(x^{3}-x_{\rm sf})}
          -{\rm e}^{\kappa_{+}(x^{3}-x_{\rm sf})}
   \right)^{2} .
\end{align}
Hereafter, we use the shorthand notation for the average over $x^{3}$ defined by
\begin{align}
      \label{eq:def-ave}
   \langle \cdots \rangle
   = \frac{ \int_{x_{\rm sf}-l_{\rm c}}^{x_{\rm sf}}{\rm d}x^{3}
            \cdots\rho^{2}}
          { \int_{x_{\rm sf}-l_{\rm c}}^{x_{\rm sf}}{\rm d}x^{3}
            \rho^{2}} .
\end{align}
The average in eq.~(\ref{eq:Lam-perp-appr}) should be identified with this.

Let us require that
${\mib n}_{\pm}$ are connected by time-reversal operation as
\begin{align}
     \label{eq:condition-n_pm}
  {\mib n}_{+}=-{\rm i}\sigma_{y}{\mib n}_{-}^{*} ,
\end{align}
indicating that if ${\mib n}_{-}=\, ^{t}(n_{1},-n_{2})$,
then ${\mib n}_{+}^{\dagger}=(n_{2},n_{1})$.
If ${\mib e}^{3}$ is parameterized in terms of
spherical coordinates $\theta(x^{1},x^{2})$ and $\phi(x^{1},x^{2})$ as
${\mib e}^{3}=(\sin\theta\cos\phi,\sin\theta\sin\phi,\cos\theta)$
and hence
\begin{align}
  \sigma^{3}
  = \left[
      \begin{array}{cc}
        \cos\theta & \sin\theta {\rm e}^{-{\rm i}\phi} \\
        \sin\theta {\rm e}^{{\rm i}\phi} & -\cos\theta
      \end{array}
    \right] ,
\end{align}
a natural solution of eq.~(\ref{eq:eigen-sigma3}) is given by
\begin{align}
        \label{dv_basis+}
  {\mib n}_{+}(x^{1},x^{2})
  & = \left[
        \begin{array}{c}
          \cos\frac{\theta}{2} {\rm e}^{-{\rm i}\frac{\phi}{2}} \\
          \sin\frac{\theta}{2} {\rm e}^{{\rm i}\frac{\phi}{2}}
        \end{array}
      \right] ,
    \\
        \label{dv_basis-}
  {\mib n}_{-}(x^{1},x^{2})
  & = \left[
        \begin{array}{c}
          \sin\frac{\theta}{2} {\rm e}^{-{\rm i}\frac{\phi}{2}} \\
          -\cos\frac{\theta}{2} {\rm e}^{{\rm i}\frac{\phi}{2}}
        \end{array}
      \right] .
\end{align}
This standard expression of ${\mib n}_{\pm}$
obviously satisfies eq.~(\ref{eq:condition-n_pm}).
Note that in the derivation of the effective surface Hamiltonian,
we require only eq.~(\ref{eq:condition-n_pm})
and do not explicitly use eqs.~(\ref{dv_basis+}) and (\ref{dv_basis-}).

Let us evaluate the matrix elements given in eq.~(\ref{eq:mat-H_eff}).
We easily find that the diagonal elements vanish,
i.e., $\langle\langle \pm|H_{\parallel}|\pm \rangle\rangle = 0$.
In evaluating the off-diagonal elements,
we should note that $\partial_{i}$ in $H_{\parallel}$ acts
not only on $|\pm \rangle\rangle(x^{1},x^{2})$
but also on $\rho(x^{3};x^{1},x^{2})$.
After straightforward calculations (see Appendix B), we find
\begin{align}
     \label{h2}
  \tilde{\cal H}_{\rm eff}
  = \left[
      \begin{array}{cc}
        0 & \tilde{\mathcal D}_{+} \\
        \tilde{\mathcal D}_{-} & 0
      \end{array}
    \right]
\end{align}
where
\begin{align}
      \label{eq:tilde-D+}
 \tilde{\mathcal D}_{+}
  & = \sum_{i=1}^{2}
      \left[
      \left(\eta_{i}A-\xi_{i}m_{2}\right)\partial_{i}
      + \frac{1}{2}\left[\partial_{i}\left(\eta_{i}A-\xi_{i}m_{2}\right)
                   \right]
      \right] ,
        \\
 \tilde{\mathcal D}_{-}
  & = \sum_{i=1}^{2}
      \left[
      - \left(\eta_{i}A-\xi_{i}m_{2}\right)^{*}\partial_{i}
      - \frac{1}{2}\left[\partial_{i}\left(\eta_{i}A-\xi_{i}m_{2}\right)^{*}
                   \right]
      \right].
\end{align}
In the above expressions for $\tilde{\mathcal D}_\pm$
we have introduced
\begin{align}
       \label{eq:eta-i}
   \eta_{i}
 & = \int_{x_{\rm sf}-l_{\rm c}}^{x_{\rm sf}}{\rm d}x^{3} \sqrt{\mathcal G}
     \rho^{2}{\mib n}_{+}^{\dagger}\sigma^{i}{\mib n}_{-}
       \nonumber \\
 & = \frac{\langle \sqrt{\mathcal G}
           {\mib n}_{+}^{\dagger}\sigma^{i}{\mib n}_{-}\rangle}
          {\langle\sqrt{\mathcal G}\rangle} ,
       \\
       \label{eq:xi-i}
   \xi_{i}
 & = 2\sum_{j=1}^{2}
     \int_{x_{\rm sf}-l_{\rm c}}^{x_{\rm sf}}{\rm d}x^{3} \sqrt{\mathcal G}
     \rho^{2}g^{ij}{\mib n}_{+}^{\dagger}\partial_{j}{\mib n}_{-}
       \nonumber \\
 & = 2\sum_{j=1}^{2}
     \frac{\langle \sqrt{\mathcal G}
           g^{ij}{\mib n}_{+}^{\dagger}\partial_{j}{\mib n}_{-}\rangle}
          {\langle\sqrt{\mathcal G}\rangle} .
\end{align}
We show in Appendix C that $\xi_{i}$ is simplified to
\begin{align}
     \label{eq:xi-simple}
  \xi_{i} = \sum_{j=1}^{2}
            \frac{\langle \sqrt{\mathcal G}{b_{j}}^{i}
                  {\mib n}_{+}^{\dagger}\sigma^{j}{\mib n}_{-}\rangle}
                 {\langle\sqrt{\mathcal G}\rangle} ,
\end{align}
where ${b_{j}}^{i} = -(\partial_{j}{\mib e}_{3})\cdot{\mib e}^{i}$.
The second term of $\tilde{\mathcal D}_{\pm}$
is essential in ensuring the hermiticity of
$\tilde{\cal H}_{\rm eff}$ when the coefficient $\eta_{i}A-\xi_{i}m_{2}$
depends on $x^{i}$.~\cite{raoux,takahashi,takane}
The effective velocity in the $x^{i}$-direction
is determined by $\eta_{i}A-\xi_{i}m_{2}$, where the second term with $m_{2}$
represents the renormalization due to curvature.~\cite{takane,imura4}

In the final step, we slightly modify the normalization of
the obtained effective Hamiltonian
so as to make it compatible with the standard convention.
Since we have inserted the factor $\sqrt{\mathcal G}$ in the definition of
the matrix elements [see eq.~(\ref{eq:mat-H_eff})],
the integral measure for the orthonormalization of $\tilde{\mib\alpha}$
is ${\rm d}x^{1}{\rm d}x^{2}$.
On the other hand,
in the 2D world onto which the electronic motion in the surface state 
is projected, the natural integral measure for the ``surface element'' is
${\rm d}\langle S \rangle
\equiv \langle\sqrt{\mathcal G}\rangle{\rm d}x^{1}{\rm d}x^{2}$.
Accordingly, we define the new two-component spinor ${\mib \alpha}$ as
\begin{align}
  {\mib \alpha} \equiv \frac{1}{\sqrt{\langle\sqrt{\mathcal G}\rangle}}
                       \tilde{\mib\alpha} ,
\end{align}
for which ${\rm d}\langle S \rangle$ can be applied.
The effective Hamiltonian for ${\mib \alpha}$ is obtained as
\begin{align}
     \label{h2-mod}
  {\cal H}_{\rm eff}
  = \left[
      \begin{array}{cc}
        0 & \mathcal{D}_{+} \\
        \mathcal{D}_{-} & 0
      \end{array}
    \right] ,
\end{align}
where
\begin{align}
    \label{44}
    {\mathcal D}_{+}
  & = \sum_{i=1}^{2}
      \biggl[
      \left(\eta_{i}A-\xi_{i}m_{2}\right)
      \left(  \partial_{i}
            + \frac{1}{2}\left[\partial_{i}
                               {\rm ln}\langle\sqrt{\mathcal G}\rangle
                         \right]
      \right)
        \nonumber \\
  & \hspace{25mm}
    + \frac{1}{2}\left[\partial_{i}\left(\eta_{i}A-\xi_{i}m_{2}\right)
                 \right]
      \biggr] ,
    \\
        \label{45}
    {\mathcal D}_{-}
  & = \sum_{i=1}^{2}
      \biggl[
      - \left(\eta_{i}A-\xi_{i}m_{2}\right)^{*}
      \left(  \partial_{i}
            + \frac{1}{2}\left[\partial_{i}
                               {\rm ln}\langle\sqrt{\mathcal G}\rangle
                         \right]
      \right)
        \nonumber \\
  & \hspace{25mm}
    - \frac{1}{2}\left[\partial_{i}\left(\eta_{i}A-\xi_{i}m_{2}\right)^{*}
                 \right]
      \biggr] .
\end{align}
In the obtained linear differential operator
the correction term,
$(1/2)\partial_{i}{\rm ln}\langle\sqrt{\mathcal G}\rangle$,
corresponds precisely to what is known as the spin connection
in the Dirac theory
on curved surfaces.~\cite{lee,zhang1,zhang2,parente,abrikosov}
Noticing that
${\rm d}\langle S \rangle
= \langle\sqrt{\mathcal G}\rangle{\rm d}x^{1}{\rm d}x^{2}$,
we have identified the precise origin of this term,
i.e., the spin connection in the {\it narrow} sense,
as arising from the spatial variations of an infinitesimal area element.
The last term in the expression for ${\mathcal D}_{\pm}$ is there,
ensuring the hermiticity of ${\cal H}_{\rm eff}$.
This term,
though having an origin different from the spin connection 
in the above narrow sense,
may be regarded, together with the previous term, 
$(1/2)\partial_{i}{\rm ln}\langle\sqrt{\mathcal G}\rangle$,
as a part of the spin connection in a {\it broad} sense.

In this section we have seen explicitly that
the low-energy electrons on the arbitrary curved surface of a topological insulator
do obey a Dirac equation.
The obtained effective Hamiltonian
takes indeed a generalized Dirac form with linear differential operators.
The set of equations,
eq.~(\ref{h2-mod}) with eqs.~(\ref{44}) and (\ref{45}),
constitutes the central result of the paper.

\section{Origin of the Berry's phase $\pi$}

Let us discuss the origin of the Berry's phase $\pi$
in the context of the boundary condition for
${\mib \alpha}(x^{1},x^{2})=\,^{t}(\alpha_{+},\alpha_{-})$,
and its relation to the so-called spin-to-surface locking.
Let us consider a situation in which either one or both of
our curvilinear coordinates $x^{i}$ are {\it cyclic}.
Angular coordinates of a closed (e.g., cylindrical or spherical) geometry
could be a typical example of such a coordinate.
In the following we assume that only the coordinate $x^{i}$ ($i=1$ or $2$)
is cyclic with a cycle of $L_{i}$
(the other coordinate does not appear explicitly in the discussion).
Our starting point is the fact that any wave function
\begin{align}
 |\Psi(x^{i})\rangle
 = \alpha_+(x^{i})|+\rangle(x^{i}) + \alpha_-(x^{i})|-\rangle(x^{i})
\end{align}
must satisfy the periodic boundary condition, i.e.,
\begin{align}
  |\Psi(x^{i})\rangle = |\Psi(x^{i}+L_{i})\rangle .
\end{align}
The local spin quantization axis ${\mib n}_{\pm}(x^{i})$
plays a crucial role in our argument.
Obviously, if ${\mib n}_{\pm}(x^{i}) = {\mib n}_{\pm}(x^{i}+L_{i})$
and hence $|\pm\rangle(x^{i}) = |\pm\rangle(x^{i}+L_{i})$,
the boundary condition for ${\mib \alpha}(x^{i})$ must be periodic as
\begin{align}
  {\mib \alpha}(x^{i}) = {\mib \alpha}(x^{i}+L_{i}) .
\end{align}
Note that ${\mib n}_{\pm}$ could change its sign as
${\mib n}_{\pm}(x^{i}) = - {\mib n}_{\pm}(x^{i}+L_{i})$
after the coordinate $x^{i}$ finishes one complete cycle of evolution
(i.e., $x^{i} \to x^{i}+L_{i}$).
If this is the case, the sign of $|\pm\rangle(x^{i})$ is also reversed as
$|\pm\rangle(x^{i}) = -|\pm\rangle(x^{i}+L_{i})$.
Accordingly, the boundary condition for ${\mib \alpha}$ must be antiperiodic as
\begin{align}
  {\mib \alpha}(x^{i}) = -{\mib \alpha}(x^{i}+L_{i}) .
\end{align}
This indicates that the boundary condition for ${\mib \alpha}$ is
simply determined by whether or not ${\mib n}_{\pm}(x^{i})$ changes its sign
when the cyclic coordinate $x^{i}$ is shifted by one complete cycle $L_i$
(in physical terms, this would correspond to one complete orbital revolution
of the Dirac electron around the closed surface).
Actually, ${\mib n}_{\pm}(x^{i})$ changes its sign
when it {\it rotates} by $\pm 2\pi$ around an arbitrary axis in the spin space
as the Dirac electron {\it revolves} once around the closed surface.

Let us observe that the antiperiodicity of the boundary condition
discussed above is equivalent to
the Berry's phase $\pi$ in the Dirac theory on curved surfaces.
In the latter point of view the antiperiodic boundary condition is
abandoned (i.e., replaced with the periodic one) 
at the cost of introducing a Berry's phase $\pi$.
This can be seen as follows: we focus on the case in which ${\mib n}_{\pm}$
changes its sign as ${\mib n}_{\pm}(x^{i}) = - {\mib n}_{\pm}(x^{i}+L_{i})$.
Then, we reformulate this problem by the use of
the following single-valued basis vectors
\begin{align}
        \label{sv_basis}
  \tilde{\mib n}_{\pm}(x_{i}) \equiv
  \exp\left({\rm i}\pi\frac{x_{i}}{L_{i}}\right){\mib n}_{\pm}(x_{i}) ,
\end{align}
which obviously satisfy the periodic boundary condition:
$\tilde{\mib n}_{\pm}(x^{i}) = \tilde{\mib n}_{\pm}(x^{i}+L_{i})$.
Reflecting the fact that eq.~(\ref{eq:condition-n_pm}) dose not hold
for $\tilde{\mib n}_{\pm}$,
the effective Hamiltonian in this single-valued basis becomes
\begin{align}
      \label{h2-sv}
  {\cal H}_{\rm eff}^{\rm sv}
  = \left[
      \begin{array}{cc}
        0 & {\mathcal D}_{+}^{\rm sv} \\
        {\mathcal D}_{-}^{\rm sv} & 0
      \end{array}
    \right] ,
\end{align}
where
\begin{align}
    {\mathcal D}_{+}^{\rm sv}
  & = \sum_{i=1}^{2}
      \biggl[
      \left(\eta_{i}A-\xi_{i}m_{2}\right)
      \left(  \partial_{i}
            + \frac{1}{2}\left[\partial_{i}
                               {\rm ln}\langle\sqrt{\mathcal G}\rangle
                         \right]
            +{\rm i}\frac{\pi}{L_{i}}
      \right)
        \nonumber \\
  & \hspace{25mm}
    + \frac{1}{2}\left[\partial_{i}\left(\eta_{i}A-\xi_{i}m_{2}\right)
                 \right]
      \biggr] ,
      \\
    {\mathcal D}_{-}^{\rm sv}
  & = \sum_{i=1}^{2}
      \biggl[
      - \left(\eta_{i}A-\xi_{i}m_{2}\right)^{*}
        \left(  \partial_{i}
              + \frac{1}{2}\left[\partial_{i}
                                 {\rm ln}\langle\sqrt{\mathcal G}\rangle
                           \right]
              + {\rm i}\frac{\pi}{L_{i}}
        \right)
          \nonumber \\
  & \hspace{25mm}
      - \frac{1}{2}\left[\partial_{i}\left(\eta_{i}A-\xi_{i}m_{2}\right)^{*}
                   \right]
      \biggr] .
\end{align}
Note that in the above formulas the change of the boundary condition
has been absorbed as a correction to derivatives,
i.e., the term of the form of ${\rm i}\pi/L_{i}$,
which sums up to a Berry's phase $\pi$.
In this sense, the Berry's phase $\pi$ is a mere rewriting of
the antiperiodicity of the basis vectors ${\mib n}_{\pm}$,
while it is often regarded as an important part of the 
spin connection in literatures.~\cite{zhang1,zhang2,imura1,spherical}

We have so far argued that the Berry's phase $\pi$ should be attributed
to the sign change of the local spin quantization axis
${\mib n}_{\pm}$ caused by a $\pm 2\pi$ rotation in the spin space.
Previously, the Berry's phase $\pi$ is interpreted as a consequence of
a $\pm 2\pi$ rotation of the real spin caused by
the spin-to-surface locking.~\cite{zhang1,zhang2,egger,bardarson,imura1}
This statement is plausible but slightly misleading in the sense explained below.
As the Dirac electron revolves once around the closed surface,
its spin inevitably rotates by $\pm 2\pi$
in the presence of the sign change of ${\mib n}_{\pm}$.
That is, the $\pm 2\pi$ spin rotation occurs
regardless of whether the spin-to-surface locking holds or not.
This indicates that the spin-to-surface locking is not essential
in the appearance of the Berry's phase $\pi$.
Indeed, even in the situation where the Berry's phase $\pi$
or equivalently the antiperiodic boundary condition plays a role,
the spin-to-surface locking does not necessarily or globally occur,
as is demonstrated in the spherical system
of a topological insulator.~\cite{spherical}

\section{Application to simple cases}

In this section we apply the general framework established so far
to the following two representative cases:
samples of either a cylindrical or a spherical shape and
find an explicit formula [corresponding to eqs.~(\ref{44}) and (\ref{45})]
of the differential operators ${\mathcal D}_{\pm}$
that specify the explicit form of the Dirac Hamiltonian~(\ref{h2-mod}).
In the analysis given below we employ orthogonal curvilinear coordinates,
for which $g_{ij}={\rm diag}(g_{11},g_{22},1)$ and
$g^{ij}={\rm diag}(g^{11},g^{22},1)$
with $g^{11}=(g_{11})^{-1}$ and $g^{22}=(g_{22})^{-1}$.

\subsection{The cylindrical case}

Let us consider an infinitely long cylindrical topological insulator
aligned along the $z$-axis with radius $R$.
We employ the following three coordinates $(x^{1},x^{2},x^{3})=(\phi, z, r)$,
in terms of which the 3D Cartesian coordinates are expressed as
$(x,y,z)=(r\cos\phi, r\sin\phi, z)$.
The parameter $x_{\rm sf}$ is simply equal to $R$.
The tangent and normal vectors are
\begin{align}
  {\mib e}_{1}
  & = (-r\sin\phi, r\cos\phi, 0) ,
       \\
  {\mib e}_{2}
  & = (0,0,1) ,
       \\
  {\mib e}_{3}
  & =(\cos\phi, \sin\phi, 0) ,
\end{align}
and
\begin{align}
  {\mib e}^{1}
  & = \left( -\frac{\sin\phi}{r}, \frac{\cos\phi}{r}, 0 \right) ,
       \\
  {\mib e}^{2}
  & = (0,0,1) ,
       \\
  {\mib e}^{3}
  & =(\cos\phi, \sin\phi, 0) .
\end{align}
The elements of the metric tensors are $g_{11}=r^{2}$ and $g_{22}=1$,
which results in ${\mathcal G}=r^{2}$,
and the coefficients of the Weingarten equation are
${b_1}^1=-r^{-1}$ and ${b_1}^2={b_2}^1={b_2}^2=0$.
From the expressions of ${\mib e}^{i}$ we obtain the spin matrices as
\begin{align}
  \sigma^{1} & = \frac{\rm i}{r}
                 \left[ \begin{array}{cc}
                           0 & -{\rm e}^{-{\rm i}\phi} \\
                           {\rm e}^{{\rm i}\phi} & 0
                        \end{array}
                 \right] ,
          \\
  \sigma^{2} & = \left[ \begin{array}{cc}
                           1 & 0 \\
                           0 & -1
                        \end{array}
                 \right] ,
          \\
  \sigma^{3} & = \left[ \begin{array}{cc}
                           0 & {\rm e}^{-{\rm i}\phi} \\
                           {\rm e}^{{\rm i}\phi} & 0
                        \end{array}
                 \right] .
\end{align}
As the unit vectors satisfying $\sigma^{3}{\mib n}_{\pm} = \pm{\mib n}_{\pm}$,
it is convenient to use those given in
eqs.~(\ref{dv_basis+}) and (\ref{dv_basis-}) at $\theta = \pi /2$,
\begin{align}
      \label{eq:n-pm-cylin}
  {\mib n}_{\pm}
    = \frac{1}{\sqrt{2}}
      \left[
        \begin{array}{c}
          {\rm e}^{-{\rm i}\frac{\phi}{2}} \\
          \pm {\rm e}^{{\rm i}\frac{\phi}{2}}
        \end{array}
      \right] .
\end{align}
Then we immediately find that
${\mib n}_{+}^{\dagger}\sigma^{1}{\mib n}_{-}={\rm i}r^{-1}$
and ${\mib n}_{+}^{\dagger}\sigma^{2}{\mib n}_{-}=1$.
Substitution of these results with $\sqrt{\mathcal G}=r$
and ${b_1}^1=-r^{-1}$ and ${b_1}^2={b_2}^1={b_2}^2=0$
into eqs.~(\ref{eq:eta-i}) and (\ref{eq:xi-simple}) yields
\begin{align}
  \eta_{1}
  & = \frac{\langle{\rm i}r^{-1}\sqrt{\mathcal G}\rangle}
           {\langle\sqrt{\mathcal G}\rangle}
    = \frac{\rm i}{\langle r \rangle} ,
        \\
  \eta_{2}
  & = 1 ,
        \\
  \xi_{1}
  & = \frac{-\langle{\rm i}r^{-2}\sqrt{\mathcal G}\rangle}
           {\langle\sqrt{\mathcal G}\rangle}
    = -\frac{\rm i}{\langle r \rangle}
       \Big\langle \frac{1}{r} \Big\rangle ,
        \\
  \xi_{2}
  & = 0 .
\end{align}
Noting that
$\partial_{\phi}{\rm ln}\langle\sqrt{\mathcal G}\rangle
=\partial_{z}{\rm ln}\langle\sqrt{\mathcal G}\rangle=0$
we finally obtain the differential operator ${\mathcal D}_{\pm}$ as
\begin{align}
  {\mathcal D}_{\pm}
  = {\rm i}\left(A+\Big\langle \frac{1}{r} \Big\rangle m_{2}\right)
    \frac{\partial_{\phi}}{\langle r \rangle}
    \pm A\partial_{z} .
\end{align}
If the penetration depth $\lambda$ for surface states is
much shorter than $R$, we can approximate as $\langle r \rangle = R$
and $\langle r^{-1} \rangle = R^{-1}$, and then ${\mathcal D}_{\pm}$
is simplified to
\begin{align}
  {\mathcal D}_{\pm}
  = {\rm i}\left(A+\frac{m_{2}}{R}\right)
    \frac{\partial_{\phi}}{R}
    \pm A\partial_{z} .
\end{align}
The effective Hamiltonian is given by eq.~(\ref{h2-mod})
with ${\mathcal D}_{\pm}$ obtained above.
The result similar to this has been reported in ref.~\citen{imura1},
where the renormalization correction $m_{2}/R$
to the effective velocity is ignored.
Let us consider the boundary condition for a spinor wave function
${\mib\alpha}(\phi,z)$.
Since ${\mib n}_{\pm}(\phi)$ in eq.~(\ref{eq:n-pm-cylin})
changes its sign when $\phi \to \phi +2\pi$,
the boundary condition for the variable $\phi$ must be antiperiodic, i.e.,
\begin{align}
  {\mib \alpha}(\phi,z) = -{\mib \alpha}(\phi+2\pi,z) .
\end{align}

\subsection{The spherical case}

We turn to the second case of a spherical topological insulator
with radius $R$.
We employ the standard spherical coordinates
$(x^{1},x^{2},x^{3})=(\theta, \phi, r)$,
in terms of which the 3D Cartesian coordinates are expressed as
$(x,y,z)=(r\sin\theta\cos\phi, r\sin\theta\sin\phi, r\cos\theta)$.
The parameter $x_{\rm sf}$ is again equal to $R$.
The tangent and normal vectors are
\begin{align}
  {\mib e}_{1}
  & =(r\cos\theta\cos\phi, r\cos\theta\sin\phi, -r\sin\theta) ,
        \\
  {\mib e}_{2}
  & =(-r\sin\theta\sin\phi, r\sin\theta\cos\phi, 0) ,
        \\
  {\mib e}_{3}
  & =(\sin\theta\cos\phi, \sin\theta\sin\phi, \cos\theta) ,
\end{align}
and
\begin{align}
  {\mib e}^{1}
  & = \left(\frac{\cos\theta\cos\phi}{r}, \frac{\cos\theta\sin\phi}{r},
              -\frac{\sin\theta}{r} \right) ,
        \\
  {\mib e}^{2}
  & = \left(-\frac{\sin\phi}{r\sin\theta}, \frac{\cos\phi}{r\sin\theta}, 0
      \right) ,
        \\
  {\mib e}^{3}
  & = (\sin\theta\cos\phi, \sin\theta\sin\phi, \cos\theta) .
\end{align}
The elements of the metric tensors are
$g_{11}=r^{2}$ and $g_{22}=r^{2}\sin^{2}\theta$,
which results in ${\mathcal G}=r^{4}\sin^{2}\theta$,
and the coefficients of the Weingarten equation are
${b_1}^1={b_2}^2=-r^{-1}$, and ${b_1}^2={b_2}^1=0$.
From the expressions of ${\mib e}^{i}$ we obtain the spin matrices as
\begin{align}
  \sigma^{1} & = \frac{1}{r}
                 \left[ \begin{array}{cc}
                           -\sin\theta & \cos\theta{\rm e}^{-{\rm i}\phi} \\
                           \cos\theta{\rm e}^{{\rm i}\phi} & \sin\theta
                        \end{array}
                 \right] ,
          \\
  \sigma^{2} & = \frac{\rm i}{r\sin\theta}
                 \left[ \begin{array}{cc}
                           0 & -{\rm e}^{-{\rm i}\phi} \\
                           {\rm e}^{{\rm i}\phi} & 0
                        \end{array}
                 \right] ,
          \\
  \sigma^{3} & = \left[ \begin{array}{cc}
                           \cos\theta & \sin\theta{\rm e}^{-{\rm i}\phi} \\
                           \sin\theta{\rm e}^{{\rm i}\phi} & -\cos\theta
                        \end{array}
                 \right] .
\end{align}
As the unit vectors satisfying $\sigma^{3}{\mib n}_{\pm} = \pm{\mib n}_{\pm}$,
it is convenient to use those given in
eqs.~(\ref{dv_basis+}) and (\ref{dv_basis-}).
We find that ${\mib n}_{+}^{\dagger}\sigma^{1}{\mib n}_{-}=-r^{-1}$
and ${\mib n}_{+}^{\dagger}\sigma^{2}{\mib n}_{-}={\rm i}(r\sin\theta)^{-1}$.
Substitution of these results with $\sqrt{\mathcal G}=r^{2}\sin\theta$,
${b_1}^1={b_2}^2=-r^{-1}$, and ${b_1}^2={b_2}^1=0$
into eqs.~(\ref{eq:eta-i}) and (\ref{eq:xi-simple}) yields
\begin{align}
  \eta_{1}
  & = \frac{-\langle r^{-1}\sqrt{\mathcal G}\rangle}
           {\langle\sqrt{\mathcal G}\rangle}
    = - \frac{\langle r \rangle}{\langle r^{2} \rangle} ,
        \\
  \eta_{2}
  & = \frac{{\rm i}\langle (r\sin\theta)^{-1}\sqrt{\mathcal G}\rangle}
           {\langle\sqrt{\mathcal G}\rangle}
    = \frac{{\rm i}\langle r \rangle}{\langle r^{2} \rangle\sin\theta} ,
        \\
  \xi_{1}
  & = \frac{\langle r^{-2}\sqrt{\mathcal G}\rangle}
           {\langle\sqrt{\mathcal G}\rangle}
    = \frac{1}{\langle r^{2} \rangle} ,
        \\
  \xi_{2}
  & = \frac{-{\rm i}\langle (r^{2}\sin\theta)^{-1}\sqrt{\mathcal G}\rangle}
           {\langle\sqrt{\mathcal G}\rangle}
    = - \frac{{\rm i}}{\langle r^{2} \rangle\sin\theta} .
\end{align}
Noting that
$\partial_{\theta}{\rm ln}\langle\sqrt{\mathcal G}\rangle=\cot\theta$ and
$\partial_{\phi}{\rm ln}\langle\sqrt{\mathcal G}\rangle=0$
we finally obtain the differential operator ${\mathcal D}_{\pm}$ as
\begin{align}
  {\mathcal D}_{\pm}
  = \left(A+\frac{m_{2}}{\langle r \rangle} \right)
    \frac{\langle r \rangle}{\langle r^{2} \rangle}
    \left(\mp \partial_{\theta}
          + {\rm i}\frac{\partial_{\phi}}{\sin\theta}
          \mp \frac{1}{2}\cot\theta
    \right) .
\end{align}
It should be emphasized that $(1/2)\cot\theta$
is identified with the spin connection
in the curved Dirac theory.~\cite{parente,abrikosov}
If the penetration depth $\lambda$ for surface states is much shorter than $R$,
we can approximate as $\langle r \rangle = R$
and $\langle r^{2} \rangle = R^{2}$, and ${\mathcal D}_{\pm}$ is simplified to
\begin{align}
  {\mathcal D}_{\pm}
  = \left(A+\frac{m_{2}}{R} \right)\frac{1}{R}
    \left(\mp \partial_{\theta}
          + {\rm i}\frac{\partial_{\phi}}{\sin\theta}
          \mp \frac{1}{2}\cot\theta
    \right) .
\end{align}
The effective Hamiltonian is given by eq.~(\ref{h2-mod})
with ${\mathcal D}_{\pm}$ obtained above.
The result similar to this has been reported in ref.~\citen{spherical},
where the renormalization correction $m_{2}/R$
to the effective velocity is ignored.
Let us consider the boundary condition for a spinor wave function
${\mib\alpha}(\theta,\phi)$.
The system is periodic with respect to $\phi$,
so we consider the boundary condition when $\phi$ is increased by $2\pi$.
Since ${\mib n}_{\pm}(\phi,\theta)$
changes its sign when $\phi \to \phi +2\pi$,
the boundary condition for the variable $\phi$ must be antiperiodic, i.e., 
\begin{align}
  {\mib \alpha}(\theta,\phi) = -{\mib \alpha}(\theta,\phi+2\pi) .
\end{align}

\section{Summary and discussion}

The behavior of low-energy electrons on an arbitrary curved surface of
3D (strong) topological insulators has been considered on general grounds.
In contrast to the specific cases studied earlier,
we have reached a unified description of such low-energy electrons
by giving the most general form of the surface Dirac Hamiltonian
[eq.~(\ref{h2-mod}) with eqs.~(\ref{44}) and (\ref{45})]
that has been explicitly derived from the bulk effective theory
in the continuum limit.
It was shown that the low-energy surface electrons do obey the Dirac equation
in this generalized form with the effective velocity renormalized by
the curved nature of the surface.
A special attention has been paid to the boundary condition for
a spinor wave function ${\mib \alpha}$,
which becomes relevant on a closed surface described by
at least one {\it cyclic} coordinate $x^{i}$.
Whether the boundary condition is periodic or antiperiodic
depends on the behavior of the local spin quantization axis ${\mib n}_{\pm}$.
If the sign of ${\mib n}_{\pm}$ is unchanged
after one complete cyclic evolution of the coordinate $x^{i}$,
the boundary condition for ${\mib \alpha}$ is periodic.
On contrary, if ${\mib n}_{\pm}$ changes its sign
due to a $\pm 2\pi$ rotation in the spin space,
the boundary condition becomes antiperiodic.
It is argued that the antiperiodicity of the boundary condition is
equivalent to the Berry's phase $\pi$.

Previously, the effective Hamiltonian for Dirac electrons
on a curved surface has been considered in a framework different from
the one presented in this paper.~\cite{lee,zhang1,zhang2,parente}
In this alternative viewpoint, one starts from
a two-dimensional Dirac equation for a flat surface,
and takes account of the curved nature of a surface
by a coordinate transformation,
resulting in the curved surface Dirac theory.
The effective Hamiltonian thus obtained contains
a fictitious vector potential called the spin connection,
which corresponds to the term
$(1/2)\partial_{i}{\rm ln}\langle\sqrt{\mathcal G}\rangle$ in our framework.
We have shown that the spin connection represents corrections arising from
the spatial variation of an infinitesimal area element.
We have also seen that such an {\it ad hoc} curved Dirac theory
overlooks the renormalization of
the effective velocity arising from the quadratic mass term $m_{2}$.
This is because the theory ignores from the outset
the three dimensional nature of the problem.
It should be noted that the above renormalization arises
even in the limit in which the penetration depth $\lambda$ of surface states
is vanishingly short, as is seen in \S5.

\section*{Acknowledgment}

The authors are supported by KAKENHI:
Y.T. by a Grant-in-Aid for Scientific Research (C) (No. 24540375)
and K.I. by the ``Topological Quantum Phenomena'' (No. 23103511).

\appendix

\section{Proof of $E_{\perp}=0$}

As noted in the text, the boundary condition
$|\psi (x^{3}=x_{\rm sf})\rangle=\mib 0$ holds only when
$|u_{+}\rangle = |u_{-}\rangle$ for $\kappa_{+} \neq \kappa_{-}$. 
Here $|u_{\pm}\rangle$ are the eigenvectors satisfying
$M_{\perp}(\kappa_{\pm})|u_{\pm}\rangle = {\mib 0}$, where
\begin{align}
  M_{\perp}(\kappa_{\pm})
  = \left[
      \begin{array}{cc}
         m_{0}-m_{2}\zeta_{\pm}-E_{\perp}
         & -{\rm i}A\kappa_{\pm}\sigma^{3} \\
         -{\rm i}A\kappa_{\pm}\sigma^{3}
         & -m_{0}+m_{2}\zeta_{\pm}-E_{\perp}
      \end{array}
    \right]
\end{align}
with $\zeta_{\pm} = \kappa_{\pm}^{2}+\Delta\kappa_{\pm}$.
It is instructive to rewrite the eigenvalue equation as
\begin{align}
    \left[
    \begin{array}{cc}
       \frac{m_{0}-m_{2}\zeta_{\pm}-E_{\perp}}{A\kappa_{\pm}}
       & -{\rm i}\sigma^{3} \\
       -{\rm i}\sigma^{3}
       & -\frac{m_{0}-m_{2}\zeta_{\pm}+E_{\perp}}{A\kappa_{\pm}}
    \end{array}
    \right] |u_{\pm}\rangle = {\mib 0} .
\end{align}
From the above equation we easily observe that
if $|u_{+}\rangle = |u_{-}\rangle$, the following two equations
\begin{align}
     \frac{m_{0}-m_{2}\zeta_{+}-E_{\perp}}{A\kappa_{+}}
 & = \frac{m_{0}-m_{2}\zeta_{-}-E_{\perp}}{A\kappa_{-}} ,
        \\
     \frac{m_{0}-m_{2}\zeta_{+}+E_{\perp}}{A\kappa_{+}}
 & = \frac{m_{0}-m_{2}\zeta_{-}+E_{\perp}}{A\kappa_{-}}
\end{align}
must hold simultaneously.
This directly results in $E_{\perp} = 0$.
Under this zero energy condition, ${\rm det}\{M_{\perp}\}=0$
yields $m_{0}-m_{2}\zeta_{\pm} = A\kappa_{\pm}$,
where $m_{0} > 0$ and $m_{2} < 0$ are assumed.
Solving $m_{0}-m_{2}\zeta_{\pm} = A\kappa_{\pm}$
with respect to $\kappa_{\pm}$, we obtain eq.~(\ref{eq:kappa-def}).

\section{Derivation of the off-diagonal matrix elements}

From eqs.~(\ref{delta_H}), (\ref{eq:pm-def}) and (\ref{eq:pm_vec-def})
it is easy to show that
\begin{align}
     \label{eq:H+-}
  \langle +| H_{\parallel} |- \rangle
  & = \int_{x_{\rm sf}-l_{\rm c}}^{x_{\rm sf}}{\rm d}x^{3}\sqrt{{\cal G}}
      \rho \biggl( A\sum_{i=1}^{2}{\mib n}_{+}^{\dagger}
                    \sigma^{i}\partial_{i}{\mib n}_{-}
     \nonumber \\
  & \hspace{25mm}
                 - m_{2}{\mib n}_{+}^{\dagger}
                   \Lambda_{\parallel}{\mib n}_{-}
           \biggr) \rho .
\end{align}
Let us denote the first and second terms in the right-hand side
of eq.~(\ref{eq:H+-}) as $\langle +| H_{\parallel} |- \rangle_{1}$
and $\langle +| H_{\parallel} |- \rangle_{2}$, respectively.
The first term is rewritten as
\begin{align}
 \langle +| H_{\parallel} |- \rangle_{1}
   = \sum_{i=1}^{2}
     \left[ \eta_{i}A\partial_{i}
          + \frac{1}{2}\left(\partial_{i}\eta_{i}A\right)
          + AP_{i}
     \right]
\end{align}
where $\eta_{i}$ is defined in eq.~(\ref{eq:eta-i}) and
\begin{align}
 P_{i}
 & = -\frac{1}{2}
      \int_{x_{\rm sf}-l_{\rm c}}^{x_{\rm sf}}{\rm d}x^{3} \rho^{2}
      \biggl[ {\mib n}_{+}^{\dagger}
              (\partial_{i}\sqrt{{\cal G}}\sigma^{i}){\mib n}_{-}
        \nonumber \\
 & \hspace{8mm}
             + \sqrt{{\cal G}}
               \Bigl((\partial_{i}{\mib n}_{+}^{\dagger})\sigma^{i}{\mib n}_{-}
                   - {\mib n}_{+}^{\dagger}\sigma^{i}(\partial_{i}{\mib n}_{-})
               \Bigr)
      \biggr] .
\end{align}
We can show $(\partial_{i}{\mib n}_{+}^{\dagger})\sigma^{i}{\mib n}_{-}
- {\mib n}_{+}^{\dagger}\sigma^{i}(\partial_{i}{\mib n}_{-})=0$
by using eq.~(\ref{eq:condition-n_pm}), and $P_{i}$ is reduced to
\begin{align}
   \label{eq:Pi-mod}
 P_{i}
   = -\frac{1}{2}
      \int_{x_{\rm sf}-l_{\rm c}}^{x_{\rm sf}}{\rm d}x^{3} \rho^{2}
      {\mib n}_{+}^{\dagger}(\partial_{i}\sqrt{{\cal G}}
      \sigma^{i}){\mib n}_{-} .
\end{align}
It is easy to show with
${\mib e}^{1} = {\mib e}_{2}\times{\mib e}_{3}/\sqrt{\mathcal G}$ and
${\mib e}^{2} = {\mib e}_{3}\times{\mib e}_{1}/\sqrt{\mathcal G}$ that
\begin{align}
  \partial_{1}\sqrt{{\cal G}}\sigma^{1}
  & = \left[ (\partial_{1}{\mib e}_{2})\times{\mib e}_{3}
            + {\mib e}_{2}\times(\partial_{1}{\mib e}_{3})
      \right]\cdot{\mib \sigma} ,
     \\
  \partial_{2}\sqrt{{\cal G}}\sigma^{2}
  & = \left[ (\partial_{2}{\mib e}_{3})\times{\mib e}_{1}
            + {\mib e}_{3}\times(\partial_{2}{\mib e}_{1})
      \right]\cdot{\mib \sigma} .
\end{align}
Applying the Gauss equation~(\ref{eq:Gauss}) and
the Weingarten equation~(\ref{eq:Weingarten}), we find that
\begin{align}
  \partial_{1}\sqrt{{\cal G}}\sigma^{1}
  & = \sqrt{\mathcal G}
      \left( \Gamma_{21}^{2}\sigma^{1}
            - \Gamma_{21}^{1}\sigma^{2}+{b_{1}}^{1}\sigma^{3}
      \right) ,
    \\
  \partial_{2}\sqrt{{\cal G}}\sigma^{2}
  & = \sqrt{\mathcal G}
      \left( - \Gamma_{12}^{2}\sigma^{1}
              + \Gamma_{12}^{1}\sigma^{2}+{b_{2}}^{2}\sigma^{3}
      \right) .
\end{align}
Substituting these into eq.~(\ref{eq:Pi-mod})
and using $\Gamma_{ij}^{k}=\Gamma_{ji}^{k}$
and ${\mib n}_{+}^{\dagger}\sigma^{3}{\mib n}_{-}=0$,
we find that $\sum_{i=1}^{2}P_{i}=0$.
Now we turn to the second term which is given by
\begin{align}
 \langle +| H_{\parallel} |- \rangle_{2}
 & = -m_{2}\sum_{i,j=1}^{2}
     \int_{x_{\rm sf}-l_{\rm c}}^{x_{\rm sf}}{\rm d}x^{3}
     \rho {\mib n}_{+}^{\dagger}
        \nonumber \\
 & \hspace{18mm} \times
     \partial_{i}\left(\sqrt{\mathcal G}g^{ij}\partial_{j} \right)
     {\mib n}_{-} \rho .
\end{align}
Since ${\mib n}_{+}^{\dagger}{\mib n}_{-}=0$,
only the terms with ${\mib n}_{+}^{\dagger}\partial_{i}{\mib n}_{-}$
or ${\mib n}_{+}^{\dagger}\partial_{i}\partial_{j}{\mib n}_{-}$
do not vanish.
It is then reduced to
\begin{align}
 \langle +| H_{\parallel} |- \rangle_{2}
 = \sum_{i=1}^{2}
   \left[ -\xi_{i}m_{2}\partial_{i}
          -\frac{1}{2}\left(\partial_{i}\xi_{i}m_{2}\right)
          +m_{2}Q_{i} \right] ,
\end{align}
where $\xi_{i}$ is defined in eq.~(\ref{eq:xi-i}) and
\begin{align}
 Q_{i}
   =  \sum_{j=1}^{2}\int_{x_{\rm sf}-l_{\rm c}}^{x_{\rm sf}}{\rm d}x^{3}
      \sqrt{\cal G}\rho^{2}g^{ij}
      (\partial_{i}{\mib n}_{+}^{\dagger})(\partial_{j}{\mib n}_{-}) .
\end{align}
We can show with eq.~(\ref{eq:condition-n_pm})
that $\sum_{i,j=1}^{2}g^{ij}
(\partial_{i}{\mib n}_{+}^{\dagger})(\partial_{j}{\mib n}_{-})=0$.
Hence $\sum_{i=1}^{2}Q_{i}=0$.
Combining the resulting $\langle +| H_{\parallel} |- \rangle_{1}$
and $\langle +| H_{\parallel} |- \rangle_{2}$
we finally arrive at eq.~(\ref{eq:tilde-D+}).
The expression of $\langle -| H_{\parallel} |+ \rangle$
can also be obtained by repeating the procedure described above.

\section{Simplification of $\xi_{i}$}

In this short Appendix we simplify the expression of $\xi_{i}$
defined in eq.~(\ref{eq:xi-i}).
The starting point is the following eigenvalue equation:
$\sigma^{3}{\mib n}_{-} = - {\mib n}_{-}$.
Differentiating this by $x^{j}$ and then constructing the inner product
between the resulting expression and ${\mib n}_{+}$, we obtain
\begin{align}
  {\mib n}_{+}^{\dagger}(\partial_{j}{\mib n}_{-})
  = - \frac{1}{2}{\mib n}_{+}^{\dagger}(\partial_{j}\sigma^{3}){\mib n}_{-} .
\end{align}
The Weingarten equation~(\ref{eq:Weingarten}) enables us to replace
$\partial_{j}\sigma^{3} = (\partial_{j}{\mib e}^{3})\cdot{\mib \sigma}$
with $-\sum_{k=1}^{2}b_{jk}\sigma^{k}$.
This results in
\begin{align}
    \label{eq:xi-app}
  \xi_{i}
   =  \sum_{j,k=1}^{2}\int_{x_{\rm sf}-l_{\rm c}}^{x_{\rm sf}}{\rm d}x^{3}
      \sqrt{\cal G}\rho^{2}g^{ij}b_{jk}
      {\mib n}_{+}^{\dagger}\sigma^{k}{\mib n}_{-} .
\end{align}
Noting that $\sum_{j=1}^{2}g^{ij}b_{jk} = {b_{k}}^{i}$,
the expression~(\ref{eq:xi-app}) for $\xi_{i}$
is reduced to eq.~(\ref{eq:xi-simple}).


\begin{thebibliography}{99}
\bibitem{spinc} 
The use of the term, ``spin connection'', in this context
may not be appropriate.
To reexamine this point is one of the main issues of the paper.
\end{thebibliography}
\end{document}